\documentclass[showpacs,preprintnumbers,amsmath,amssymb]{revtex4}
\usepackage{graphicx}
\usepackage{dcolumn}
\makeatletter
\input{epsf}

\newcommand{\bd}{\begin{document}}
\newcommand{\ed}{\end{document}}
\newcommand{\bc}{\begin{center}}
\newcommand{\ec}{\end{center}}
\newcommand{\bfr}{\begin{flushright}}
\newcommand{\efr}{\end{flushright}}
\newcommand{\vs}{\vspace}
\newcommand{\hs}{\hspace}
\newcommand{\beq}{\begin{equation}}
\newcommand{\eeq}{\end{equation}}
\newcommand{\lb}{\linebreak}
\newcommand{\mb}{\makebox}
\newcommand{\fb}{\framebox}
\newcommand{\mc}{\multicolumn}
\newcommand{\ben}{\begin{enumerate}}
\newcommand{\een}{\end{enumerate}}
\newcommand{\bit}{\begin{itemize}}
\newcommand{\eit}{\end{itemize}}
\newcommand{\un}{\underline}
\newcommand{\lefq}{\lefteqn}
\newcommand{\ba}{\begin{array}}
\newcommand{\ea}{\end{array}}
\newcommand{\beqa}{\begin{eqnarray}}
\newcommand{\eeqa}{\end{eqnarray}}
\newcommand{\beqas}{\begin{eqnarray*}}
\newcommand{\eeqas}{\end{eqnarray*}}
\newcommand{\bfg}{\begin{figure}}
\newcommand{\efg}{\end{figure}}
\newcommand{\bds}{\begin{displaymath}}
\newcommand{\eds}{\end{displaymath}}
\newcommand{\btb}{\begin{tabbing}}
\newcommand{\etb}{\end{tabbing}}
\newcommand{\para}{\parallel}
\newcommand{\pad}{\partial}
\newcommand{\nn}{\nonumber}
\newcommand{\la}{\leftarrow}
\newcommand{\ra}{\rightarrow}
\newcommand{\lgla}{\longleftarrow}
\newcommand{\lgra}{\longrightarrow}
\newcommand{\La}{\Leftarrow}
\newcommand{\Ra}{\Rightarrow}
\newcommand{\Lra}{\Leftrightarrow}
\newcommand{\Lgla}{\Longleftarrow}
\newcommand{\Lgra}{\Longrightarrow}
\newcommand{\bm}{\boldmath}
\newcommand{\lan}{\langle}
\newcommand{\ran}{\rangle}
\renewcommand{\a}{\alpha}
\renewcommand{\b}{\beta}
\newcommand{\g}{\gamma}
\newcommand{\G}{\Gamma}
\renewcommand{\d}{\delta}
\newcommand{\eps}{\epsilon}
\newcommand{\s}{\sigma}
\newcommand{\lam}{\lambda}
\newcommand{\D}{\Delta}
\newcommand{\vare}{\varepsilon}
\newcommand{\pr}{\prime}
\newcommand{\ro}{\rho}
\newcommand{\nab}{\nabla}
\newcommand{\m}{\mu}
\newcommand{\n}{\nu}
\newcommand{\Sg}{\Sigma}
\newcommand{\p}{\pi}
\newcommand{\R}{I\!\!R}
\newcommand{\om}{\omega}
\newcommand{\Om}{\Omega}
\newcommand{\ze}{\zeta}
\newcommand{\vart}{\vartheta}
\newcommand{\tri}{\triangle}
\newcommand{\f}{\frac}
\newcommand{\iny}{\infty}
\newcommand{\pro}{\propto}
\begin{document}
\title{Quantum Hall Effect In Bilayer Systems And
The Noncommutative Plane: A Toy Model Approach}
\author{B. Basu}
\email{banasri@isical.ac.in}
\author{Subir Ghosh}
 \email{sghosh@isical.ac.in}
\affiliation{Physics and Applied Mathematics Unit\\
 Indian Statistical Institute\\
 Kolkata-700108 }

\begin{abstract}
We have presented a quantum mechanical toy model for the study of
Coulomb interactions in Quantum Hall (QH) system. Inclusion of
Coulomb interaction is essential for the study of {\it{bilayer}}
QH system and our model can simulate it, in the compound state, in
a perturbative framework. We also show that in the noncommutative
plane, the Coulomb interaction is modified at a higher order in
the noncommutativity parameter $\theta$, and only if $\theta$
varies from layer to layer in the QH system.
\end{abstract}

\maketitle

Physical conditions for the occurrence (as well as experimental
observation) of Integer Quantum Hall Effect (IQHE)\cite{ez1}
allows us to reduce the many electron system to an effective
single electron one. This miraculous simplification is possible
due to the incompressibility of the planar electron gas in the
presence of a perpendicular high magnetic field at very low
temperature and without disorder \cite{ez1}. The incompressibility
of the ground state forbids the creation of separated
particle-hole pairs, (capable of carrying current), because their
creation requires energy. This is the key point of the quantum
Hall system which at the same time suggests the usage of single
particle picture. It can be explicitly  shown that  quantum
mechanics of a single electron in a magnetic field is able to
capture the discrete nature of Hall conductivity. Indeed, this
simple model can not address most of the details of QH physics,
notably among them formation of the Hall plateauxs, the (passive)
role of disorder and impurities etc. (see for example Stone in
\cite{ez1}).

Compared to  IQHE, which is essentially a one body effect,
explanation of Fractional Quantum Hall Effect (FQHE) \cite{ez1}
requires the introduction of many body physics since the latter is
induced by inter-particle forces,  Coulomb interactions (between
electrons confined to the two-dimensional plane) being the
dominant one. The  magnetic field can be large enough to
effectively freeze out both the kinetic and spin Zeeman energies.
In \cite{la} an explanation of the $\nu=\frac{1} {2n+1}$ FQH
states was given. In an attractive alternative approach, quantum
Hall effect was also studied in spherical geometry \cite{hal}
which was pursued \cite{bb1} in the Berry phase framework. This
formalism can nicely \cite{bb2} accommodate the newly observed
\cite{pan} FQH states within the primary sequence. The concept of
composite particles play an important role in understanding the
mechanism of incompressibility in  FQHE. This last fact motivates
us to question how far one can pursue the study of a planar many
body interacting system in the QH regime, in terms of single
particle states the describe IQHE. An interesting idea, mooted in
\cite{jel}, is that FQHE might be interpreted as IQHE in the
Non-Commutative (NC) plane, where the NC parameter $\theta $
generates the fractional QH states. In the present Letter, we
first elaborate our proposed toy model for a bilayer QH system and
subsequently consider the effects of the NC coordinates onit.
 Hopefully the method and results
presented here can be generalized to study a restricted class of
systems exhibiting FQHE.

Coulomb interactions are essential in the study of bilayer QH
systems: a new variety of QH states.\cite{ez2} A bilayer system is
made by trapping electrons in two thin layers at the interface of
semiconductors. There exists a variety of QH states depending on
the relative strength of various interactions. The key
interactions are the {\it{inter-layer}} Coulomb interaction
(~$e^2/4\pi \epsilon \sqrt{r^2+d^2}$), the {\it{intra-layer}}
Coulomb interaction (~$e^2/4\pi \epsilon r$), where $r$ is the
planar inter-particle distance and $d$ is the inter-layer
distance. Typical Coulomb energy scales are $e^2/4\pi \epsilon d$
and $e^2/4\pi \epsilon l$ where $l=\f{1}{\sqrt{eB}}$ denotes the
magnetic length (in natural units $c=\hbar=1$). We are neglecting
the Zeeman interaction  and the tunnelling interaction. Our
analysis is relevant in the context of the simplest of bilayer
states, the {\it{compound}} state where charges are not
transferable between the two layers. Addition of the monolayer
Hall conductivities yields the bilayer Hall conductivity. For this
reason, we consider the states of bilayer system as product states
of two monolayer single particle states. In this setup we will
compute the Coulomb energy in the lowest order of perturbation
theory. Qualitative agreement between our observations and recent
rigorous field theoretic  results \cite{ez2} indicate that we are
on the right track.

Recent excitement in the High Energy physics community regarding
the possibility of having a Non-Commutative (NC) spacetime (or
space) \cite{rev} has triggered the study of effects of NC space
in different systems (for a review of NC effects on planar
theories, see \cite{sg}). In \cite{jel} it was shown that NC
effects renormalize the magnetic field $B$ and in turn the filling
factor $\nu=2\pi\rho /\sqrt{eB}$, where $\rho$ denotes the
electron density. This induces a modification in the Hall
conductivity $\sigma $ which is proportional to $\nu$. We will
consider these NC effects on our toy model of the bilayer system.

Let us start by considering a monolayer QH system, the (Lowest
Landau Level) states of which are reproduced by the effective
single particle  states of a planar charged  spinless particle in
a magnetic field $B$. The Hamiltonian of the  model, in a crossed
background electric field ($E$) is given by,
\begin{equation}\label{h1}
H=\frac{1}{2m}\left[\left(p_{x}-\frac{eB}{2c}y\right)^2~+~
\left(p_{y}-\frac{eB}{2c}x\right)^2\right]+~eEx~,
\end{equation}
where the magnetic field is expressed in the symmetric gauge
\cite{jel}
$${\bf{B}}=\nabla \times {\bf{A}},~~A_x=-\frac{B}{2}y, ~A_y=\frac{B}{2}x$$  and $m$ and $e$
denote the mass and charge of the particle. (Our notations and conventions agree
with \cite{jel}.) In order to solve $H$, we change
the variables to
\begin{equation}
z=x+iy~~~~~, p_z=\frac{1}{2}(p_x-ip_y)
\end{equation}
and introduce two sets of creation and annihilation operators:
$$
b^\dagger=-2ip_{\bar z}+\frac{eB}{2c}z+\lambda ~;~~
b=-2ip_z+\frac{eB}{2c}{\bar z}+\lambda~,$$
\begin{equation}
d^\dagger=-2ip_{\bar z}-\frac{eB}{2c}z~;~~
d=2ip_z-\frac{eB}{2c}{\bar z}
\end{equation}
where $\lambda=\frac{mcE}{B}$.

These two sets commute with each other and satisfy the commutation
relations
\begin{equation}
[b,b^\dagger]=2m\omega~,~~ [d,d^\dagger]=-2m\omega
\end{equation}
where $\omega=\frac{eB}{mc}$ is the cyclotron frequency. The
Hamiltonian can be rewritten as
\begin{equation}
H=\frac{1}{4m}(b^\dagger
b+~bb^\dagger)-\frac{\lambda}{2m}(d^\dagger~+d)-\frac{\lambda^2}{2m}
\end{equation}
As expected, the electric field lifts the degeneracy. The
eigenfunctions and the energy spectrum of the Hamiltonian $H$ are
$$
\psi_{(n,\alpha)}=\phi_n \otimes \phi_\alpha= |n,\alpha >,$$
\begin{equation}\label{wave}
\phi_n =\frac{1}{\sqrt{(2m\omega)^n n!}}(b^\dagger)^n|0>~;~~
\phi_\alpha =\exp\left(i(\alpha y+i\frac{m\omega}{2}xy)\right),
\end{equation}
\begin{equation}
E_{(n,\a)}= \frac{ \omega}{2}(2n+1)-~\frac{
\lambda}{m}\alpha-\frac{\lambda^2}{2m},~~~~n=0,1,2....,~~\alpha\in
R .
\end{equation}

The generic form of Hall current, induced by the Lorentz force, is
\begin{equation}
J_i=\frac{e^2\nu}{2\pi}
 \epsilon_{ij}E_j
 \end{equation}
where $\nu=2\pi l^2 \rho$ is the filling factor with $\rho$ the
 electron density. In the present case, the current is derived from the equation of motion:
\begin{equation}\label{c1}
J_i=ie\rho [H,r_i]=\frac{e\rho}{m}(p_i +\frac{e}{c}A_i).
\end{equation}
The expectation value of $J_i$ in the eigen-states
$\psi_{n,\alpha}$ in (\ref{wave}) is,
\begin{equation}
<J_x>=0~,~~<J_y>=-\frac{ec\rho}{B}E~. \label{j2}
\end{equation}
Hence one can read off the Hall conductivity $\sigma$ as,
\begin{equation}
\sigma=2\pi e^2\nu. \label{j3}
\end{equation}
For a system with $n$ filled Landau levels, one gets
\begin{equation}
\sigma=n(2\pi e^2\nu). \label{jn3}
\end{equation}
Although the correct quantization of the Hall conductance has been
achieved, quite naturally this simplistic model is unable to
address a host of phenomena related to QH effect, most notably
among them the formation of Hall plateaux, (absence of) effects of
disorders and impurities etc..

Our idea is to carry through this effective single particle
picture as far as feasible in the realm of bilayer QH systems. We
will model the bilayer QH system in the compound state as a weakly
coupled system of two mono-layers with negligible tunnelling
between them ($\sim $ large $d$). We take the compound state
Hamiltonian ($H_C$) as,
\begin{equation}
H_{C}=H_1(x_1,y_1) + H_2(x_2,y_2),
\end{equation}
where $H(x,y)$ is given in (\ref{h1}). This is just a combination
of two decoupled mono-layers. Obviously the wave functions of the
compound system will be of the form, \beq \Psi_C=[\phi^1_{n_1}
\otimes \phi^1_{a_1}] \otimes [\phi^2_{n_2} \otimes \phi^2_{a_2}]
~. \eeq This type of direct product states have appeared before in
\cite{ajel} in the context of   electron-hole pair in a quantum
Hall system, without Coulomb interaction. The Hall current
$J_{(C)i}$ for the compound state is given by,
\begin{equation}
J_{(C)i}=ie\rho [H_C,r_{(1)i}+r_{(2)i}]. \label{jc}
\end{equation}
Clearly the Hall conductivity for the Compound system is obtained
as,
\begin{equation}
\sigma_{C}=(n_1+n_2)\frac{e^2}{\nu}, \label{j33}
\end{equation}
where $n_1$ and $n_2$ comes from the two mono-layers.

After this somewhat trivial rederivation of the bilayer Compound
state QH conductivity  we come to the interesting part: effect of
the Coulomb interaction. We intend to confirm that the Coulomb
interaction will not affect the conductivity in the present
scenario. Subsequently we will compute the correction in energy at
the lowest non-trivial order of perturbation. Quite surprisingly,
our "back of the envelope" estimate agrees with the rigorous
findings \cite{ez2}, at least structurally.

Conventionally, the Coulomb interaction in bilayer systems is
decomposed  in to two parts \cite{ez1,ez2},
\begin{equation}\label{c2}
H_{Coulomb}=\Delta H_1 +\Delta H_2~;~~
 \Delta H_1= \frac{C_1e^2}{4\pi\epsilon_0}\frac{1}{|{\bf x}|}~,~~
\Delta H_2= \frac{C_2e^2}{4\pi\epsilon_0}\frac{1}{\sqrt{|{\bf
x}|^2+d^2}} ,
\end{equation}
where $|{\bf x}| =[(x_1-x_2)^2+(y_1-y_2)^2]^{1/2}$. Clearly
$\Delta H_1$ and $\Delta H_2$ represent monolayer and bilayer
contributions respectively. $C_1$ and $C_2$ are two numerical
parameters which can introduce screening effects since the charges
are not in vacuum. However, we will not elaborate on them any
further since we only wish to present the structure of the energy
correction, and not absolute numerical values.

Before proceeding any further, the expression of Coulomb term in
the Hamiltonian, as posited in (\ref{c2}), requires an
explanation. We have borrowed this term from its (more rigorous)
field theoretic counterpart which reads \cite{sec9},
 \begin{equation}\label{ft1}
 H_{Coulomb}=\frac{1}{2}\sum_{\alpha,\beta}\int d^2xd^2y V_{\alpha\beta}(\bf{x} -\bf {y})\rho^{\alpha}(\bf{x})\rho^{\beta}(\bf{y}).
 \end{equation}
 Here $\alpha$ and $\beta$ denote the top or bottom layer and $\rho^{\alpha}$
 refers to the corresponding charge density. The potential is given by,
 \begin{equation}\label{ft2}
V_{\alpha\beta}({\bf x} -{\bf y})=\frac{e^2}{4\pi\epsilon}\frac{1}{\sqrt{\mid {\bf x}-{\bf
 y}\mid^{2}+d^{2}_{\alpha\beta}}},
\end{equation}
with $d_{tt}=d_{bb}=0$ and $d_{tb}=d$, the inter-layer separation.
Notice that this pseudospin framework, (where the electrons in top
and bottom layer are distinguished by "up" or "down" pseudospin),
admits the possibility of tunnelling of electrons between the
layers, apart from their static interactions. As we mentioned in
the beginning, in our toy model approach the tunnelling of
electrons between layers is not considered. Since we concentrate
on the static energy, the bilayer nature  of the system is
manifested only through $d$. This is as if the bilayer system is
projected on to a single layer. The intra and inter-layer Coulomb
terms differ by the appearance of $d$ which introduces a lower
bound in the inter particle distance in the inter-layer Coulomb
effect.

Indeed, deduction of the first result is easy. The definition of
the Hall current (\ref{c1}) shows that $J_C$ will remain unaltered
since the position operators will commute with the Coulomb
interaction terms (\ref{c2}).

Next we compute the energy correction,
\begin{equation}
\Delta E_C= <\psi_C|\Delta H_1|\psi_C> + <\psi_C|\Delta
H_2|\psi_C>, \label{en}
\end{equation}
in the first  perturbative order  only. In terms of the creation
and annihilation operators we express the coordinates as,
 \beq
x_{\alpha}=\frac{1}{2eB}
(b_{\alpha}+b_{\alpha}^\dagger-d_{\alpha}-d_{\alpha}^\dagger-2\lambda),~~
y_{\alpha}= -\frac{i}{2eB}
(b_{\alpha}^\dagger-b_{\alpha}-(d_{\alpha}^\dagger-d_{\alpha}),
\eeq where $\alpha =1,2 $ indicate the two mono-layers. The
inter-particle distance can now be expressed in terms of the
ladder operators:
 $$ {\bf {x}}^2=
(x_1-x_2)^2+(y_1-y_2)^2 $$ \beq =\f{1}{e^2B^2}[b_1^\dagger b_1+
b_2^\dagger b_2+\f{1}{2}\{(d_1
+d_2^\dagger)^2-(d_1-d_2^\dagger)^2\}+2m\omega + D], \eeq where
$$D=b_1^\dagger d_2 + d_1^\dagger b_2 + b_1 d_2^\dagger +d_1 b_2^\dagger -
 b_1^\dagger b_2- b_1 b_2^\dagger-  d_1^\dagger d_2- d_1 d_2^\dagger -  b_1^\dagger d_1 -
 d_1^\dagger b_1-b_2^\dagger d_2-d_2^\dagger b_2$$

A simple calculation yields
\begin{eqnarray} \label{c3}
<\psi_C| \Delta H_1|\psi_C>= \f{C_1e^2}{4\pi\epsilon_0}eB
<\psi_C|\f{1}{\left[{2m\omega(2n+1)+2\a ^2+(\f{eB}{c})^2(y_1^2+y_2^2)+
D}\right]^{1/2}}|\psi_C> \end{eqnarray}
 Let us first put forward our results.
For large $n$ the above relation can be approximated to yield,
 \beq
 <\psi_C\mid\Delta H_1|\psi_C> \cong \f{C_1e^2}{4\pi\epsilon_0}eB\f{1}{\sqrt{A}}
 (1-\f{\a^2}{2A}-\f{2\pi l^2e^2B^2}{3A})
 \eeq
 where $A=2m\omega(2n+1)\cong 4nm\omega $. Notice that in the first order of
 perturbation $<\psi_C|D|\psi_C>=0$. The other point to note is that $y$-operator
 matrix elements are integrated to give $2\pi l^2$, the effective area of confinement
 of the electrons. The above result is simplified to give,
  \beq
<\psi_C\mid \Delta H_1|\psi_C>\cong
\f{C_1e^2}{4\pi\epsilon_0}[\f{1}{2\sqrt{n}}\f{1}{l}
    -\f{\a^2}{8n}l-\f{1}{24n\sqrt{n}}\f{2\pi l^2}{l^3}].
\eeq In a similar way, expanding in powers of $\f{1}{d}$, we find,
 \beq <\psi_C\mid
\Delta H_2|\psi_C>\cong
\f{C_2e^2}{4\pi\epsilon_0}[\f{1}{d}-\f{\a^2l^4}{2d^3}-\f{2\pi
l^2}{3} \f{1}{d^3}].\eeq Collecting all the term, we finally
obtain,
  $$ \Delta E_C =<\psi_C\mid
\Delta H_1+ \Delta H_2\mid\psi_C>$$ \beq \cong
N\f{e^2}{4\pi\epsilon_0}[C_1\{\f{1}{2\sqrt{n}}\f{1}{l}
    -\f{\a^2}{8n}l-\f{1}{24n\sqrt{n}}\f{1}{l^3}2\pi l^2\}
+C_2\{\f{1}{d}-\f{\a^2l^4}{2d^3}-\f{2\pi l^2}{3d^3}\} ], \eeq
where $N$ indicates the total number of electrons in the system.
The system energy is proportional to $N$ since we have exploited
the (non-interacting) quasi-particle picture.

Before proceeding further, a  comment regarding our large $n$
assumption in (\ref{c3}) is pertinent. It may appear odd that our
result, instead of being valid for small $n$ near ground state, is
better suited for large $n$. However, one has to remember that it
is not simply the Coulomb energy in a two particle system that we
are after. Quite obviously the Landau level wavefunctions that we
have exploited are not the natural ones for such an analysis. On
the contrary, our aim is to simulate the behavior of charges in a
QH (many body) system, subjected to Coulomb interaction. In fact,
our large $n$ restriction seems to be in order if we keep in mind
the fact that large $n$ means the oscillator states are more
localized and there is less of overlap. This is in agreement with
our model of the weakly coupled bilayer QH system having product
wavefunctions in the compound state.

For $\alpha =0$ and large $n$, the leading terms in $\Delta E_C$
are, \beq \label{c4}\Delta E_C \cong
\f{e^2N}{4\pi\epsilon_0}[C_1(\f{1}{2\sqrt{n}}\f{1}{l})
+C_2(\f{1}{d}-\f{2\pi l^2}{3d^3}) ]. \eeq Let us try to compare
our findings with the recent field theoretic computations of
 the Coulomb energy in bilayer quantum Hall state, as obtained by Ezawa et.al.\cite{ez2}.
 The Coulomb energy in ground state, for large $\frac{d}{\sqrt{2}l}$, is
 (see equations (3.4)-(3.8) of \cite{ez2}),
  \beq \label{c5}
E_C\cong
\frac{e^2}{4\pi\epsilon_0}\frac{N}{4}[\frac{1}{l}+\frac{1}{d}-\frac{l^2}{d^3}].
\eeq Comparing with (\ref{c5}) we find that in our result
(\ref{c4}), the relevant parameters, {\it{i.e.}} $N,~l$ and $d$
have appeared correctly in our expression with proper signs. This
concludes our analysis of the IQHE  in a quantum mechanical toy
model describing the  bilayer system. \vskip .5cm \noindent
{\bf{Bilayer QH system in noncommutative space:}}\\
\vskip .5cm \noindent Let us now analyze what happens if the
noncommutativity in the space coordinates is switched on. In
particular, we wish to study how it affects the Coulomb
interaction. The NC ($\tilde x,\tilde y)$ plane obey the following
algebra: \beq [\tilde x,\tilde y]=i\theta. \eeq A representation
of this structure in terms of commuting $(x,y)$ coordinates  is,
\begin{equation}
\tilde x\equiv x-\frac{\theta}{2}p_y ~,~~\tilde y\equiv
y+\frac{\theta}{2}p_x . \label{nc}
\end{equation}
This means that one can replace the monolayer Hamiltonian $H$ by
its NC analogue $\tilde H$, \beq {H_{nc}}=
\frac{1}{2m}\left[\left( (1-\kappa)p_x-\frac{eB}{2}y\right)^2
~+~\left(
(1-\kappa)p_y-\frac{eB}{2}x\right)^2\right]~+~eE\left(x-\frac{\theta}{2}p_y\right)
\eeq where $\kappa=\frac{e\theta B}{4}$. Diagonalization of
$\tilde H$ yields (see \cite{jel} for details) the energy
spectrum, \beq \tilde E=\frac{ \tilde{\omega}}{2}(2n+1)- \gamma
\lambda_{+}\a-\frac{m}{2}\lambda_{-}^2 \eeq with \beq
\lambda_{\pm}= \frac{mcE}{B}~\pm~\frac{emE\theta}{4\gamma} \eeq
with $\tilde{\omega}=\gamma \omega$ and $\gamma=1-\kappa$. For
reasons that will become apparent later, we will consider a
generalization of the above form of noncommutativity: \beq [\tilde
x_1,\tilde y_1]=i\theta_1~,~~ [\tilde x_2,\tilde y_2]=i\theta_2.
\eeq Thus we are taking $\theta$ to be different for the two
planes. This is in effect a restricted form of non-constant or
space dependent $\theta$. This form of $\theta$ has appeared in
the literature. In our model, the energy for the compound state
will be, \beq \tilde E_C=\frac {\omega}{2}[(2n_1+1)\gamma
_1+(2n_2+1)\gamma _2]- \a[\gamma _1 \lambda_{1+}+\gamma _2
\lambda_{2+}]-\frac{m}{2}[\lambda_{1-}^2 +\lambda_{2-}^2].\eeq The
notation in the above is self-explanatory.

It is straightforward to consider the NC effects on the energy due
to the Coulomb interaction term. Instead of that, let us study the
effect of the NC-Coulomb term on conductivity. Remember that the
Coulomb interaction had no effect on the conductivity in the
commutative plane. In the present case, $\Delta H_1$ will remain
unchanged since it refers to the same plane and depends on the
relative position. However, $\Delta H_2$ will be modified since it
pertains to two different planes and will depend on
\begin{eqnarray}
[d^2+(\tilde x_1-\tilde x_2)^2+(\tilde y_1-\tilde y_2)^2]&
=&[d^2+\{ x_1-
x_2-\frac{1}{2}(\theta_1{p_1}_y-\theta_2{p_2}_y)\}^2\nonumber\\&+&\{
y_1- y_2 +\frac{1}{2}(\theta_1{p_1}_x-\theta_2{p_2}_x)\}^2].
\end{eqnarray} Thus NC effect in the conductivity operator in the
compound state will be given by, \begin{eqnarray} [\Delta
H_{nc},(x_1+x_2)]&= &\frac{i C_2 e^2}{8\pi\epsilon}
(\theta_1-\theta_2) [d^2+(\tilde x_1-\tilde x_2)^2+(\tilde
y_1-\tilde y_2)^2]^{-3/2}\nonumber \\
&\times&(\theta_1{p_1}_x-\theta_2 {p_2}_x)
\end{eqnarray} and
\begin{eqnarray} [\Delta H_{nc},(y_1+y_2)]&= &\frac{i C_2
e^2}{8\pi\epsilon}(\theta_1-\theta_2) [d^2+(\tilde x_1-\tilde
x_2)^2+(\tilde
y_1-\tilde y_2)^2]^{-3/2} \nonumber\\
&\times&(\theta_1{p_1}_y-\theta_2 {p_2}_y)
\end{eqnarray}
This operator correction is of higher order in $\theta$. But what
is more interesting is that in this case the  QH effect is lost
since both the Hall currents are present. The expectation value of
these correction terms are not reproduced here. Since this effect
depends on $\theta_1-\theta_2$ it will vanish for
$\theta_1=\theta_2$.

If $\theta_1=\theta_2$ then from the above equations we can say
that there is no change in the conductivity. But if the two planes
have different parameters of noncommutativity then there is some
change. Possibility of having different values of $\theta$ in a
single system has been considered in \cite{ajel}.

To conclude, in the present work, we have constructed a quantum
mechanical toy model that can capture the effects of Coulomb
interaction in a quantum Hall system. Our model simulates a
bilayer quantum Hall system in the compound state and our results,
(such as energy spectrum),  agrees qualitatively with those of
Ezawa et.al. \cite{ez2}. The Coulomb term does not affect the Hall
conductivity. Following the ideas of Dayi and Jellal \cite{jel},
that fractional QH effect might be interpreted as integer QH
effect in a noncommutative plane, we extend our bilayer model in
noncommutative  coordinates. We show that the noncommutative
extension of the Coulomb term differs from its commutative
counterpart in a higher order in $\theta$ and that too only if
$\theta$ varies from layer to layer. As a future work, our aim is
to extend our model to include spin effects in to account. \vskip
.5cm {\bf Acknowledgement:} We thank the Referee for the
constructive comments.



\end{document}